\documentclass{aa}
\usepackage{graphicx}
\begin{document}
   \title{Stellar evolution with rotation and magnetic fields:}

   \subtitle{I. The relative importance of rotational and magnetic effects} 

\author{Andr\'e Maeder, Georges Meynet}

     \institute{Geneva Observatory CH--1290 Sauverny, Switzerland\\
              email:  andre.maeder@obs.unige.ch\\
              email: georges.meynet@obs.unige.ch }

   \date{Received  / Accepted }

   \offprints{Andr\'e Maeder} 

   \abstract{We compare the current effects of rotation in stellar evolution to those of
   the magnetic field created by the Tayler instability. In stellar
   regions, where magnetic field can be generated by the dynamo due to differential
   rotation (Spruit \cite{Spruit02}), we find that the growth rate
    of the magnetic instability is much faster
   than for the  thermal instability. Thus, meridional circulation is negligible with 
   respect  to the  magnetic fields, both for the transport of
   angular momentum and of chemical elements. Also, the horizontal coupling by
   the magnetic  field, which reaches values of a few $10^5$ G, is much more important than the effects  
   of the horizontal turbulence. The field, however, is not sufficient to distort the shape
   of the  equipotentials. We impose the condition  that the energy of the magnetic field
    created by the Tayler--Spruit
   dynamo cannot be larger than  the energy excess present in the differential
   rotation.  This leads to a  criterion for the existence of the magnetic field
   in stellar interiors.
   
   Numerical tests are made in a rotating star model of 15 M$_{\odot}$ rotating with
   an initial velocity of 300 km$\cdot$s$^{-1}$. We find that the coefficients  of diffusion for the transport 
   of angular momentum by the magnetic field  are several orders of magnitude
   larger than the transport coefficients for meridional circulation and shear mixing.
   The same applies to the diffusion coefficients for the chemical elements, however
   very close to the core, the strong $\mu$--gradient reduces the mixing by the magnetic instability
   to values not too different from the case without magnetic field. We also find that magnetic instability
   is present throughout the radiative envelope, with the exception of the very outer
   layers, where differential rotation is insufficient to build the field, a fact consistent 
   with the lack of evidence of strong fields at the surface of massive stars.

               }
   \maketitle
%

\section{Introduction}

The inclusion of new physical effects in stellar evolution greatly improves the
comparisons with observations. About  twenty  years ago the impact of mass loss
by stellar winds on the evolution was found to be large. However, some significant discrepancies
remained and the inclusion of rotation has enabled substantial progress
in the comparisons with observed chemical abundances, with number counts and
with chemical evolution of galaxies (cf. Langer et al. \cite{langeragb};
Maeder \& Meynet \cite{MMaraa}).
Magnetic field is coming next, but certainly 
not the last, in this series of effects which may influence all the 
outputs of stellar evolution. 

In this work, we  focus mainly on the relative importance of the effects
of the magnetic field and of rotational instabilities to try to determine which effects can be let aside
and what must be considered as a priority. Sect. 2 summarizes the main effects of the
magnetic field we are considering here following Spruit (\cite{Spruit99}, \cite{Spruit02}).
 Sect. 3 compares
the characteristic times  of meridional circulation and of magnetic field
instabilities. Sect. 4 considers what happens to the horizontal turbulence 
in presence of magnetic fields. Sect. 5 shows a new physical limit on the occurrence
of the magnetic field
in rotating stars. Sect. 6 gives some numerical values on the size of magnetic and 
rotational effects in the
case of a 15 M$_{\odot}$ star. Sect. 7 presents the conclusions.

\section{Basic properties of the magnetic field}

Let us collect here some basic expressions and concepts we need below.
Spruit (\cite{Spruit99}, \cite{Spruit02}) has shown that the magnetic field
can be created in radiative layers of stars in differential rotation.
Even  a small toroidal field is subject to an instability (called Tayler instability
by Spruit), which creates a vertical field component. Differential rotation winds up
this vertical component,  so that many new  horizontal field lines are produced. These
horizontal field lines become progressively closer and denser 
in a star in a state of differential rotation and therefore 
a much stronger  horizontal field is built. This is the dynamo processs described by Spruit.
Tayler instability is a pinch--type instability. As shown by
Spruit (\cite{Spruit99}), it has a very low threshold and is characterized by a short 
timescale. Also it  is the first instability to occur. The magnetic shear instability 
may  be present, but it is of much less importance (Spruit \cite{Spruit99}).

The instability occurs in radiative zones and two cases are considered by Spruit (\cite{Spruit99},    
\cite{Spruit02}) depending on the thermal and $\mu$--gradients through the associated 
oscillation frequencies,

\begin{equation}
N_{\mathrm{T}}^{2}\; = \; \frac{g \delta}{H_{\mathrm{P}}}(\nabla-\nabla_{\mathrm{ad}})
\label{NT}
\end{equation}
\noindent
and
\begin{equation}
N_{\mu}^2 \; = \; \frac{g \varphi}{H_{\mathrm{P}}} \nabla_{\mu}  \; .
\label{Nmu}
\end{equation}

\noindent
The thermodynamic coefficients $\delta$ and $\varphi$ are defined as follows
$\delta= - \left(\frac{\partial \ln \rho}{\partial \ln T}\right)_{\mathrm{P},\mu}$ and
$\varphi=  \left(\frac{\partial \ln \rho}{\partial \ln \mu}\right)_{\mathrm{T,P}}$.
The quantity $H_{\mathrm{P}}$ is the local pressure scale height.
Following Spruit, we call case 0 the case  where the $\mu$--gradient dominates
over the thermal gradient, i.e. when $N_{\mu} > N_{\mathrm{T}}$.
Case 1 applies  when the thermal gradient is the main restoring force, i.e.
when $N_{\mu} < N_{\mathrm{T}}$.
Let us point out that cases 0 and 1 are the two simpler  limits of a more general
case, where a proper account of both thermal and $\mu$ effects would be made.
 We may wonder whether we miss some important physical situations with this
simplification. Indeed, numerical tests show that in practice the intermediate
situation between cases 0 and 1 introduced by Spruit concerns a small but
significant part of the star as shown in Fig.~\ref{fn}.
In this part, cases 0 and 1 give diffusion coefficients which have
the same order of magnitude, therefore the general effect is correctly described.
 However, there is little doubt that in future 
the general non-adiabatic case has to be examined in detail.

The growth rate of the magnetic instability is 

\begin{equation}
\sigma = \; \frac{\omega_{\mathrm{A}}^2}{\Omega} \; ,
\label{sigma}
\end{equation}

\noindent
where $\omega_{\mathrm{A}}$ is the Alfve\'n frequency, i.e. the frequency
of magnetic waves. As shown by Pitts  \& Tayler (\cite{Pitts86}), the
reduction factor $\frac{\omega_{\mathrm{A}}}{\Omega}$ in 
Eq.~(\ref{sigma}) results from the Coriolis
force in a star rotating with angular velocity $\Omega$. The Alfv\'en frequency is

\begin{equation}
\omega_{\mathrm{A}} = \frac{B}{(4 \pi \rho)^{\frac{1}{2}} \; r} \; .
\end{equation}

\noindent
The magnetic instability works only if the unstable displacements do
not loose too much energy against the stable stratification. For this
to be the case, the radial displacements (against the buoyancy force)
must be small compared to the horizontal displacements. Taking $r$
as a maximum for the horizontal displacements, this sets an upper
limit on the radial length scale $l_{\mathrm{r0, 1}} $ of the displacements,

\begin{equation}
l_{\mathrm{r0}} < \; \frac{r \; \omega_{\mathrm{A}}}{N_{\mathrm{\mu}}}  \; ,
\label{max0}
\end{equation}
\noindent
and
\begin{equation}
l_{\mathrm{r1}} < \; r \; \left(\frac{\Omega}{N_{\mathrm{T}}}\right)^{\frac{1}{2}} 
  \left(\frac{K}{r^2 \; \Omega}\right)^{\frac{1}{4}} \; ,
 \label{max1}
\end{equation}

\noindent
where the indices 0 or 1 refer to the two cases considered. The term
$K$ is the thermal diffusivity 
$K= \frac{4ac T^3}{3 \kappa \rho^2 C_{\mathrm{p}}}$, where the other 
quantities have their usual meaning in stellar structure.
There is also a minimal extent of the magnetic instability,
below which it is quickly dissipated by magnetic diffusivity,

\begin{equation}
l_{\mathrm{r 0, 1}}^2 >  \; \frac{\eta \; \Omega}{\omega_{\mathrm{A}}^2}  \; ,
\label{min}
\end{equation}

\noindent
where $\eta$ is the diffusivity of the magnetic field.
The combination of Eq.~(\ref{min}) with Eqs.~(\ref{max0})  and (\ref{max1}) leads
to a minimum value of the Alfv\'en frequency  for the occurrence of the magnetic 
instability in the two cases considered,
\begin{equation}
\left(\frac{\omega_{\mathrm{A}}}{\Omega}\right)_0 > \; 
\left(\frac{N_{\mu}}{\Omega}\right)^{\frac{1}{2}} \left(\frac{\eta}{r^2 \; \Omega}\right)^{\frac{1}{4}} \; ,
\label{cond0}
\end{equation}

\begin{equation}
\left(\frac{\omega_{\mathrm{A}}}{\Omega}\right)_1 > \; 
\left(\frac{N_{\mathrm{T}}}{\Omega}\right)^{\frac{1}{2}} \left(\frac{K}{r^2 \; \Omega}\right)^{\frac{1}{4}} 
\left(\frac{\eta}{K}\right)^{\frac{1}{2}}  \; .
\label{cond1}
\end{equation}

\noindent
These are the conditions in order that the magnetic field overcomes the
restoring  force of buoyancy. In addition in the second case, the effects of the 
thermal diffusivity described by $K$ are also  accounted for. The corresponding 
maximum radial dimensions of the instabilities are given by the above
Eqs.~(\ref{max0}, \ref{max1}).

Spruit (\cite{Spruit02}; Eqs.~18, 19)
considers the amplification timescale necessary to double the component 
$B_{\varphi}$ starting from the radial component $B_{\mathrm{r}}$  over the largest
characteristic lengths defined by Eqs.~(\ref{max0}, \ref{max1}). He assumes the equality
of the amplification timescale
with the timescale for the damping by magnetic diffusivity over the above 
lenghtscales. In this way, he obtains
the expressions for the 
Alfv\'en frequency. In the first case, where $N_{\mu}$ dominates, this is

\begin{equation}
\left(\frac{\omega_{\mathrm{A}}}{\Omega}\right)_0  = \; q\; \frac{\Omega}{N_{\mathrm{\mu}}} \; ,
\label{omegaA0}
\end{equation}

\noindent
where  $q= -\frac{\partial \ln \Omega}{\partial \ln r}$. Thus, we
see that the Alfv\'en frequency  (a measure of  the field amplitude) 
depends on the differential rotation parameter and on the ratio
of the angular velocity to the Brunt--V\H{a}is\H{a}l\H{a} frequency.
When thermal diffusion is accounted for ($N_{\mu}$ negligible), one has

\begin{equation}
\left(\frac{\omega_{\mathrm{A}}}{\Omega}\right)_1  = \; q^{\frac{1}{2}} 
 \left(\frac{\Omega}{N_{\mathrm{T}}}\right)^{\frac{1}{8}} 
  \left(\frac{K}{r^2 \; N_{\mathrm{T}}}\right)^{\frac{1}{8}} \; ,
  \label{omegaA1}
  \end{equation}

\noindent
there the thermal diffusivity also intervenes. A few other recalls will be made when
necessary.

\section{Growth rate of the magnetic instability with respect to meridional
circulation}

A major question arises concerning  a rotating star with a magnetic field.
 What happens to the meridional circulation in
presence of the magnetic field  of the Tayler--Spruit dynamo? Basically, meridional
circulation occurs because  thermal  equilibrium cannot be achieved on
an equipotential inside a rotating star. Thus, we may wonder whether the
horizontal breakdown of  thermal equilibrium in a rotating star 
can be compensated  by a magnetic stress 
on an equipotential.

Let us define a velocity $U_{\mathrm{magn}}$ characterizing  the radial growth of the magnetic 
instability. In the two cases 0 and 1, we   consider the ratio
 of the appropriate maximum lengths given by 
Eqs.~(\ref{max0}, \ref{max1})  to the characteristic time $\sigma^{-1}$,

\begin{equation}
U_{\mathrm{magn, \; 0, 1}}= \; l_{\mathrm{r \; 0, 1}} \; \sigma  \; .
\label{U}
\end{equation}

\noindent
For the case 0 with  $N_{\mu} > N_{\mathrm{T}}$, one
gets 

\begin{equation}
U_{\mathrm{magn, \; 0}}= \; \frac{r \; \omega_{\mathrm{A}}^3}{N_{\mu} \; \Omega}  \; .
\label{U0prep}
\end{equation}

\noindent
Then, using the above expression (Eq.~\ref{omegaA0}) for the Alfv\'en frequency, we obtain

\begin{equation}
U_{\mathrm{magn, \; 0}}= \; q^3\; r \; \Omega  \left(\frac{\Omega}{N_{\mu}}\right)^4 \; .
\label{U0}
\end {equation}

\noindent
The magnetic instability grows very fastly with the angular velocity and the differential rotation parameter $q$.
In case 1 with
$N_{\mu} <  N_{\mathrm{T}}$, we get from Eq.~(\ref{U}) with
Eqs.~(\ref{max1}) and (\ref{sigma})

\begin{equation}
U_{\mathrm{magn, \;1}}= \; r \; \left(\frac{\Omega}{N_{\mathrm{T}}}\right)^{\frac{1}{2}} 
  \left(\frac{K}{r^2 \; \Omega}\right)^{\frac{1}{4}} \; 
  \frac{\omega_{\mathrm{A}}^2}{\Omega} \; .
  \label{U1prep}
  \end{equation}
 
  \noindent
  Using  Eq.~(\ref{omegaA1}) for the Alfv\'en frequency,
  one has for the growth velocity of the magnetic instability
  
  \begin{equation}
  U_{\mathrm{magn, \;1}}= \; q \; r\; \Omega 
 \left(\frac{\Omega}{N_{\mathrm{T}}}\right)^{\frac{1}{2}} 
  \left(\frac{K}{r^2 \; N_{\mathrm{T}}}\right)^{\frac{1}{2}} \; .
  \label{U1}
  \end{equation}
  
  \noindent
  This velocity also grows with $\Omega$ and $q$, but less than in the case 0,
  because  the radiative diffusivity reduces the dependence of the
  Alfv\'en frequency  with respect to rotation parameters $\Omega$ and $q$.
  
  These velocities have to be compared  with the radial component
  of the radial part of the velocity of
  the meridional circulation $U(r)$, as given by Maeder \& Zahn (\cite{MZ98};
  Eq.~4.38). If the circulation velocity would largely dominates, this would mean that the magnetic
  field has effects which are  too weak   to influence the meridional
  circulation and the circulation would develop as usually supposed.
  If on the contrary, one has $U_{\mathrm{magn, \;0, 1}} >> U(r)$, this means
  that the magnetic instability develops much faster than the thermal instability 
  at the origin of meridional circulation. 
  If so, this means  that the thermal instability created by rotation on an equipotential
  will  interact firstly with the magnetic field. Tayler
   instability, which has the shortest timescale, may possibly develop 
   and create a magnetic field which will
  introduce some  stress horizontally on the equipotential.
  Detailed    calculations must  be done  with  
   accounting  for the effects of the magnetic field in the equation for the
   entropy conservation at the basis for the calculations  of meridional
   circulation, (this has been made  for the effects of horizontal turbulence
   on the meridional circulation
   by Maeder \& Zahn \cite {MZ98}).
   
   For now, we assume  in this 
   case, in the whole region where $U_{\mathrm{magn, \;0, 1}} >> U(r)$,
   that the usual circulation velocity must be set to zero. 
   This working hypothesis is largely confirmed by the comparison below (cf. Tables 1 and 2)
   of  the velocities $U(r)$ 
   of meridional circulation to  the velocities
   $U_{\mathrm{magn,0, 1}}$ characterizing  the growth of the magnetic instability, the 
   last ones being  4 to 7 orders of a magnitude larger than the first one.

Before looking more to the numerical values, we note that
some caution has to be taken so that the comparisons of
 $U_{\mathrm{magn, \;0, 1}}$ and  $U(r)$ are done in a 
 consistent way. The expression for the transfer of angular momentum
 by circulation and diffusion $D$ is (see Zahn \cite{Zahn92}; 
 Maeder \& Zahn \cite {MZ98})

\begin{eqnarray}
\frac{d}{d t} \left(\rho r^2 \overline{\Omega}\right)_{M_r} =
\frac{1}{5 r^2}  \frac{\partial}{\partial r} \left(\rho r^4 \overline{\Omega}
  U(r)  \right)\nonumber \\[2mm]
 \qquad + \frac{1}{r^2} \frac{\partial}{\partial r}
\left(\rho \; D \; r^4 \frac{\partial\overline{\Omega}}{\partial r} \right) \; ,
\label{angmomentum}
\end{eqnarray} 

\noindent
where $\overline{\Omega}(r)$ is the average angular velocity on the equipotential.
From the second member of this expression, we see that to any diffusion coefficient
$D$,  one can define an associated velocity $U_{\mathrm{D}}$
\begin{equation}
U_{\mathrm{D}} = 5 \; \frac{D}{\Omega} \; \frac{\partial \Omega}{\partial r}= 
\; 5 \frac{D}{r} \; q \; .
\label{U5}
\end{equation}

\noindent
The factor 5 results from the integration of the angular momentum over 
the star,  which leads to  the above Eq.~(\ref{angmomentum}).
  In order to be correct, the comparison of velocities 
must be made over the corresponding quantities as they appear in Eq.~(\ref{angmomentum})
for the angular momentum transfer, 
which in turn implies Eq.~(\ref{U5}). Now, we take the  coefficients of magnetic diffusion  $\eta_{\mathrm{0}}$
and $\eta_{\mathrm{1}}$, which are obtained by replacing the inequalities of 
Eqs.~(\ref{cond0}, \ref{cond1}) by equalities. This means (cf. Spruit \cite{Spruit02})
that we consider the cases of marginal stability. Now, we can associate 
integrated velocities to these diffusion coefficients thanks to Eq.~(\ref{U5})
and we find that they are just equal to the velocities $U_{\mathrm{magn, \;0}}$ and
$U_{\mathrm{magn, \;1}}$ given above multiplied by a factor of 5. These  velocities characteristic
 of the magnetic instability given by Eq.~(\ref{U5})  are  those to be compared to the velocity of
meridional circulation, as it appears in the equation for the transport of angular momentum.
As we shall see below, even if we omit these various considerations
about  Eq.~(\ref{angmomentum}), which result  in the above  factor of 5,  the conclusions
on the order of magnitude will be the same.

Below in Sect. 6, 
a star model with an initial mass of 15 M$_{\odot}$, a composition given by
$X=0.705$ and $Z=0.02$ and an initial rotation  velocity of 300 km$\cdot$s$^{-1}$ has  been
computed. The prescriptions for rotation are those in Maeder \& Meynet (\cite{MMVII}).
Tables 1  and 2 show the main parameters when the central H--content is $X_{\mathrm{c}}=0.60$
at an age of $4.17 \cdot 10^6$ yr.
We  see that the velocities $5 \;U_{\mathrm{magn, \;0, 1}}$ are $3 \cdot 10^4$ to
$3 \cdot 10^7$ times larger that the velocity $U(r)$ of meridional circulation.
This shows that, even if present, the velocity of  meridional circulation
is negligible with respect to the corresponding velocity  characterizing the
transport of angular momentum by the magnetic field. Thus the question
whether or not meridional circulation appears in the presence of magnetic
field is of little relevance, since meridional circulation would anyway be 
totally negligible  in comparison of the effect of magnetic instability.
The  same remark arises when one compares the magnetic 
   diffusion coefficients for the angular momentum as given in 
   Eqs.~(\ref{Dang0}, \ref{Dang1}, \ref{Dang1P}) with the corresponding expression 
   (\ref{Dcirc}) for meridional circulation.

Thus, we suggest that in the  stellar regions where magnetic field is present,
{\emph {and only there}} (see Sect. 5), the meridional circulation may be neglected
with respect to magnetic field effects for the transport of angular momentum.
As is shown in Sect. 4 below, the above conclusion also applies to the transport of the 
chemical elements.
 
\section{The importance of magnetic stress vs. horizontal turbulence}

\subsection{Horizontal coupling}

The turbulence in rotating stars is highly anisotropic and has a strong horizontal component,
described by a diffusion coefficient $D_{\mathrm{h}}$,
 because  vertically the  thermal  gradient stabilizes turbulence.  This horizontal turbulence 
 strongly reduces the horizontal differential rotation, so that rotation varies only
 radially. The rotation is therefore said to be ``shellular'' (Zahn \cite{Zahn92}, $\Omega$ uniform
 at the surface of isobaric shells).
 The coefficient $D_{\mathrm{h}}$ also plays a role in the mixing of the chemical elements and 
 meridional circulation (Maeder \& Zahn \cite{MZ98}). A first expression for $D_{\mathrm{h}}$
 was given by Zahn (\cite{Zahn92}). Recently another expression of this coefficient
 has been obtained (Maeder \cite{MD_h}),

\begin{eqnarray}
D_{\mathrm{h}} =  A \; r \; \left[r
\overline{\Omega}(r) \; V \;
 \left( 2 V - \alpha U \right)\right]^\frac{1}{3} \; ,
\label{nuh}
\end{eqnarray}

\noindent with $A \approx 0.10$,  $V(r)$ the horizontal component of meridional circulation and $\alpha= \frac{1}{2}
\frac{d \ln r^2 \Omega}{d \ln r}$. Typically, this new estimate of $D_{\mathrm{h}}$ is of the order of
$10^{11}$ to 10$^{12}$ cm$^2 \cdot$s$^{-1}$in a massive star, i.e. typically $10^2$ to $10^3$
times larger than the coefficient previously estimated. Recent developments by Zahn (private
communication) and entered into model calculations by Palacios (private communication) confirm this higher values of $D_{\mathrm{h}}$.

The question is now: what happens to this horizontal turbulence, even if it is much larger 
than initially supposed,
in the presence of the magnetic field? According to Spruit (\cite{Spruit02}),
the Tayler instability and the associated dynamo leads to horizontal field
components in cases 0 and 1,

\begin{equation}
B_{\varphi, \mathrm{0}} = (4 \;\pi \; \rho)^{\frac{1}{2}} r \;q  \;\frac{\Omega^2}{N_\mu} \; ,
\label{Bphi0}
\end{equation}

\begin{equation}
B_{\varphi, \mathrm{1}} = (4 \;\pi \; \rho)^{\frac{1}{2}} r \; \Omega \; q^{\frac{1}{2}}
\left(\frac{\Omega}{N_{\mathrm T}} \right)^{\frac{1}{8}}
\left(\frac{K}{r^2 N_{\mathrm T}} \right)^{\frac{1}{8}}  \;.
\label{Bphi1}
\end{equation}

\noindent
For the numerical model of Tables 1 and 2, we find an horizontal field
\begin{equation}
B_{\varphi, \mathrm{0}}= 4.35 \cdot 10^5\  \rm{G} \quad \rm{at \;} M_{r}/M_\odot=6.57 \; ,
\end{equation}
 
 \begin{equation}
B_{\varphi, \mathrm{1}}= 2.80 \cdot 10^5\  \rm{G} \quad \rm{at \;} M_{r}/M_\odot=11.04 \; .
\end{equation}

\noindent
These fields are high with respect to our current standards,
but the magnetic pressure,

\begin{equation}
P_{\mathrm{magn}}= \frac{B^2}{8 \; \pi} 
\end{equation}

\noindent
is small with respect to the total pressure. One has typically a ratio 
$P_{\mathrm{magn}}/P= 1.5 \cdot 10^{-6}, \; 2.7 \cdot 10^{-6}$ in the above
examples. Thus, the magnetic field generated by the
Tayler--Spruit mechanism does not affect the shape of the equipotentials.
The ratios $B_{\mathrm{r}}/B_{\varphi}$ of the radial to the horizontal field
 components are given by  Eqs. (21) and (23) by Spruit (\cite{Spruit02}) and
 the values are of the order of $10^{-3}$ to a few $10^{-2}$.
 
 Now, the question is how does the horizontal coupling due to the magnetic field 
 $B_{\varphi, \mathrm{0, 1}}$ compares with the horizontal turbulence characterized
 by $D_{\mathrm{h}}$. Spruit ({\cite{Spruit02}; Eqs. 31, 32}) has given  an estimate of the coefficient
of the vertical coupling by  magnetic field. However, we cannot use  it here,
 because the field is very anisotropic.  The 
 coefficient $D_{\mathrm{B}_{\mathrm{h}}}$ for the horizontal coupling must be much larger
 than the vertical one. At low rotation, 
  we suggest to take the following estimate,
  
  \begin{equation}
  D_{\mathrm{B}_{\mathrm{h}}} \approx \;r^2 \; \omega_{\mathrm{A}} \approx
   \frac{r \; B_{\varphi}}{ (4 \pi \;\rho)^{\frac{1}{2}}}  \;.
  \end{equation}
 \noindent
 At high rotation, the growth rate is also reduced by the Coriolis
 factor $\omega_{\mathrm{A}}/\Omega$ (as suggested by Spruit in a private communication),
 thus one has then
 
  \begin{equation}
  D_{\mathrm{B}_{\mathrm{h}}} \approx \;r^2 \; (\omega_{\mathrm{A}}^2/\Omega) \approx
   \frac{ B_{\varphi}^2}{ 4 \pi \;\rho \; \Omega}  \;.
  \end{equation}

\noindent
Interestingly enough, this coefficient is of the order of   $ 10^{14}$ 
 cm$^2 \cdot$s$^{-1}$
in the above examples. Thus,  the horizontal magnetic coupling is 
 about $10^3$ larger than the horizontal coupling by turbulence,
 even when we take the value given by the new
coefficient $D_{\mathrm{h}}$ in Eq.(\ref{nuh})  above. Therefore, our conclusion
is that in the presence of Tayler--Spruit dynamo,  we can anyway neglect the 
coupling  due to horizontal turbulence with respect to the coupling insured by 
the horizontal component of the magnetic field. More likely, we 
suspect that the horizontal turbulence may not develop in this case or occur in a very 
different and limited way.

\subsection{Remarks on further consequences: the case of $D_{\mathrm{eff}}$} 

In models with rotation and without magnetic fields, the combined effect
of meridional circulation and horizontal turbulence may be treated
 as a diffusion with a coefficient usually called
 $D_{\mathrm{eff}}$  (cf. Chaboyer \& Zahn \cite{Chab92}),
 
 \begin{equation}
 D_{\mathrm{eff}}= \frac{\left[r\; U(r)\right]^2}{30 \; D_{\mathrm{h}}} \;.
 \label{Deff}
 \end{equation}
 
 \noindent
 Now, since $D_{\mathrm{h}}$ is negligible, we may wonder what happens
 to the transport of chemical elements by meridional circulation. 
 As well known, it is in general not correct to express an advection
 as a diffusion, but let us do it here just for the purpose of comparing
 orders of magnitude. The diffusion coefficient we may associate
 to the meridional circulation is in this case,
 
 \begin{equation}
 D_{\mathrm{circ}} \approx |r \; U(r)| \; .
 \label{Dcirc}
 \end{equation}
 
\noindent
This must be compared to the coefficients  $D_{\mathrm{chem,\; 0,\; 1}}$,
which are obtained by imposing  equality in Eqs.~(\ref{cond0}, \ref{cond1})
and taking $D_{\mathrm{chem,\; 0,\; 1}}= \eta_{\mathrm{\; 0,\;1}}$. The equality
means that one considers the case of marginal stability  as determining for
the transport.
 Then, in the obtained 
relations, the Alfv\'en frequency is expressed with the help
of Eqs.~(\ref{omegaA0}, \ref{omegaA1}). In this way, Spruit obtains 
(\cite{Spruit02}; Eqs.~42, 43),
 
\begin{equation}
D_{\mathrm{chem,\; 0}}= \; r^2\; \Omega\; q^4 \left(\frac{\Omega}{N_{\mu}}\right)^6 \; ,
\end{equation}
\noindent
and
 \begin{equation}
D_{\mathrm{chem,\; 1}}= \; r^2\; \Omega\; q \left(\frac{\Omega}{N_{\mathrm{T}}}\right)^{\frac{3}{4}}
\left(\frac{K}{r^2 \; N_{\mathrm{T}}}\right)^{\frac{3}{4}} \; .
\end{equation}

\noindent
In the numerical example of Tables 1 and 2, we find   ratios
$\frac{D_{\mathrm{circ}}}{D_{\mathrm{chem}}} = 0.11, \; 3.2 \cdot 10^{-5},\;
1.9 \cdot 10^{-4}$ at the 3 layers considered, (taking in each case the appropriate
expressions for the case 0 and 1). Thus, we see that through most
of the star the circulation (if it exists!) would have a negligible effect for the transport
of chemical elements. Only close to the
core, the circulation
might represent about 10 \% of the transport by the magnetic instability.
This is small anyway and owing to the profound doubts expressed above on the
occurrence
of any circulation, we consider that any transport of chemical elements by 
meridional circulation  can generally  be neglected in the presence of 
Tayler--Spruit dynamo. 

\section{An energy condition for the magnetic field}

\subsection{Energy condition}
 In a radiative zone, the magnetic field created by the Tayler--Spruit instability arises
 from differential rotation. 
 Therefore, we must impose the condition that {\emph{the
 energy in the magnetic field cannot be larger than the excess energy in differential
 rotation}}. Tayler (\cite{Tay73}) and   Pitts and Tayler (\cite{Pitts86}) have
  studied the conditions for the appearance of the magnetic instabilities, 
  which have very short growth times. The above condition is different, it is an
   energy condition, which must be satisfied on longer timescales.
   Due to the magnetic diffusivity, the field once created tends to disappear.
   Taking the values of $D_{\mathrm{chem}}= \eta$
   given for example in Fig.~6 (see also Table 1 and 2), which are between $10^{10.5}$ and $10^{12}$
    cm$^2\cdot$s$^{-1}$ over  most of the star except very  close to the convective 
    core, we find that the timescale  $\tau \approx \frac{R^2}{\eta}$
    for the diffusion of the magnetic field is of the order of 
    2 $ \cdot 10^3$ to 8 $ \cdot 10^4$ yr., which is short with respect to the
    stellar lifetimes. This means that the field created by Tayler--Spruit dynamo
    will exist only if the  energy of the magnetic field is
    continuously replenished by differential rotation during the MS evolution. Therefore, we
    need to apply the above condition.

 For a magnetic instability with a displacement 
 of amplitude $\xi$, the kinetic energy $ E_{\mathrm{B}}$
 by unit of mass is 
 
 \begin{equation}
 E_{\mathrm{B}}= \frac{1}{2} \; \omega_{\mathrm{A}}^2  \; \xi^2  \; .
 \label{Emagn}
 \end{equation}
 
 \noindent
 The excess energy $E_{\Omega}$ in the differential rotation is the difference of energy
 between the existing flow with differential rotation and a flow with an average rotation over the 
 considered radial distance $r$. Let us consider 
 a horizontal velocity field  $W(r)$. Over a vertical distance $dr$, the 
 energy excess $dE_{\Omega}$ over the extent of the magnetic instability  is 
 
 \begin{equation}
 dE_{\Omega}=  \frac{1}{2} \left[ W^2 + (W+dW)^2 \right] -\frac{1}{2} \cdot 2
 \left(W+\frac{dW}{2}\right)^2 =
 \frac{dW^2}{4} \; .
 \label{EB}
 \end{equation}
 
 \noindent
 If $l_{\mathrm{r}}$ and $l_{\mathrm{h}}$ are the 
radial and horizontal components of the displacement $\xi$ of the magnetic oscillation
 (cf. Spruit \cite{Spruit02}), the energy excess over the displacement $\xi$
 can be written,
 
 \begin{equation}
 E_{\Omega}= \frac{1}{4} \left(\frac{dW}{dr}\right)^2 \xi^2 \;
 (\frac{l_{\mathrm{r}}}{l_{\mathrm{h}}})^2   \; .
 \label{Eomega}
 \end{equation}

\noindent
The horizontal velocity
$W(r)= r \sin \vartheta \; \Omega$, where $\vartheta$ is the  colatitude. 
Thus, one has

\begin{equation}
\frac{dW}{dr} = r \sin \vartheta  \; \frac{d \Omega}{dr}=
\Omega  \sin \vartheta \; \frac{d \ln\Omega}{d \ln r}=
- \Omega \sin \vartheta \; q
\label{dudr}
\end{equation}

\noindent 
The condition  $E_{\Omega} > E_{\mathrm{B}}$ leads to

\begin{equation}
\frac{1}{2} \Omega^2 q^2 \sin^2 \vartheta  \; (\frac{l_{\mathrm{r}}}{l_{\mathrm{h}}})^2 > \omega_{\mathrm{A}}^2 \; .
\label{critprelim}
\end{equation}

This would apply to a given colatitude $\vartheta$. This equation
as it stands  means that the reservoir of
available rotational energy is larger at the equator, while we know (cf. Spruit
\cite{Spruit99}) on the other side  that Tayler instability is stronger away from equator.
 Now, for the ratio $(\frac{l_{\mathrm{r}}}{l_{\mathrm{h}}})$
of the vertical to the horizontal displacement, we could take for example in the outer 
layers, where differential rotation is generally small, the value given by Eq.~(\ref{max1})
above (case 1). This promptly leads to the following criterion 
with account of Eq.~(\ref{omegaA1}),

\begin{equation}
|q| > 3 \; \left(\frac{N_{\mathrm{T}}}{\Omega}\right)^{\frac{1}{4}}
\left(\frac{r^2 N_{\mathrm{T}}}{K}\right)^{\frac{1}{4}}  \; .
\label{false}
\end{equation}

\noindent
This equation ignores the geometry of the field.
However, we have to consider carefully 
the geometry of the problem in 2 specific respects:\\
1.-- We must  account for the 
fact that in the model of shellular rotation
the energy of rotation is not a local quantity depending on
$(r, \vartheta)$, but on $r$ only. The physical reason in usual rotating models is
the strong horizontal turbulence (cf. Zahn \cite{Zahn92}). In the present models, 
the horizontal magnetic coupling is even stronger as seen above in Sect. 4.1, so that 
shellular rotation is a valid assumption here.  In such a case, 
the average stellar structure of the rotating star
very well corresponds to the structure at a colatitude $\vartheta$ given by
 the root of the second 
Legendre polynomial $P_{2}(\cos \vartheta) =0$, (this has been verified in a recent work, Maeder \cite{Mdr02}). Thus it is appropriate to consider for the differential rotation $\frac{dW}{dr}$
in Eq.(\ref{dudr}) on a given  equipotential
the average value   $\left(\frac{dW}{dr}\right)^2 = \frac{2}{3} \; \Omega^2 q^2$. \\
--2. The geometry of the field is also particular (cf. Spruit \cite{Spruit99,Spruit02}).
It consists in stacks of magnetic loops concentric with the rotation axis.
The main component of the displacement due to Tayler instability 
is perpendicular to the rotation axis. This means that at colatitude 
$\vartheta$, the ratio $(\frac{l_{\mathrm{r}}}{l_{\mathrm{h}}})$ behaves 
essentially as $\tan \vartheta$. Since the polar caps are most unstable, while the 
equatorial regions are not,  it is thus
clear that in polar regions one has  a ratio 
$(\frac{l_{\mathrm{r}}}{l_{\mathrm{h}}})$  smaller than 1.
However, in 1--D models as here we must consider the significant average
 for the whole range of colatitudes.  The field behaves as
$B_{\varphi} \approx \sin \vartheta \cos \vartheta$ 
(cf. Spruit \cite{Spruit99}; Eq.35) and the Tayler instability develops only for
$\vartheta \leq  \pi/4$, therefore
it is necessary  on a given isobar to consider  colatitudes  $\vartheta$ smaller or
at most equal to $\pi/4$.  If we take this last value as the limit, this gives the upper
bound $\tan \vartheta = (\frac{l_{\mathrm{r}}}{l_{\mathrm{h}}}) \leq  1 $.

From these two geometrical remarks, we obtain
the necessary condition for the existence 
of a magnetic field generated by the Tayler--Spruit dynamo as follows,
 
\begin{equation}
\frac{\omega_{\mathrm{A}}}{\Omega} \; < \; \frac{|q|}{\sqrt{3}}  \; .
\label{critgen}
\end{equation}

\noindent
It means that the degree of differential rotation $q$ must at least be larger 
than $\sqrt{3}$ times the ratio $\frac{\omega_{\mathrm{A}}}{\Omega}$
of the Alfv\'en to the rotation frequency, in order that there is enough
energy in the differential rotation to allow  the Tayler--Spruit dynamo
to operate and build a magnetic field. We insist that this is only a necessary condition.
If this condition is not realized, there
is certainly no magnetic field created by the dynamo. Further works may perhaps
lead to an even more constraining condition.
The numerical factor, here $\sqrt{3}$,
may  depend on  the exact geometry of the magnetic displacements in a rotating star.
We note also that
if there are several types of instabilities  generated by differential
rotation, the available energy given by Eq.~(\ref{Eomega}) would in some way be 
shared between the instabilities. However, as mentioned above, the Tayler instability is 
the main one and  shear instabilities appear negligible in comparison.

We can go a step further, since the Alfv\'en frequency $\omega_{\mathrm{A}}$
is a function of rotation and differential parameter $|q|$.
For the case 0, where the $\mu$--gradient dominates,
 the ratio $\frac{\omega_{\mathrm{A}}}{\Omega}$
is given by the above Eq.~(\ref{omegaA0}). Thus,the above condition
(\ref{critgen}) becomes in this case

\begin{equation}
\frac{\Omega}{N_{\mu}} \; < \; \frac{1}{\sqrt{3}} = 0.5774  \ .
\label{crit0}
\end{equation}

\noindent
This is the necessary condition in order that a magnetic field may develop
from differential rotation in  case 0. At first glance, this condition
may look strange, since it means that $\Omega$ must be smaller that some value.
 The reason is that $\omega_{\mathrm{A}}$
grows like $q \; \frac{\Omega^2}{N_\mu}$, while the upper limiting $\omega_{\mathrm{A}}$
expressed by Eq.~(\ref{critgen})
goes like $\frac{|q|}{\sqrt{3}} \; \Omega$. Thus, if $\Omega$
would be too big, the actual $\omega_{\mathrm{A}}$ would  overcome the critical value.
In the numerical examples  of Tables 1 and 2,
 case 0 is relevant at the edge of the core 
at $M_{\mathrm{r}}/M_{\odot}= 6.57$. There $\frac{\Omega}{N_\mu}$
is 0.11, (typically $\frac{\Omega}{N_\mu}$ lies between 0.1 and 0.15
throughout the star). This is smaller than 
$0.5774$ and thus  the magnetic field can be present in these layers.

We now consider case 1 with thermal diffusion, the Alfv\'en frequency
is given by Eq.~(\ref{omegaA1}). With the above condition
(\ref{critgen}), one obtains

\begin{equation}
\frac{|q|}{3} \; > \; \left(\frac{\Omega}{N_{\mathrm{T}}}\right)^{\frac{1}{4}}
\left(\frac{K}{r^2 \; N_{\mathrm{T}}}\right)^{\frac{1}{4}}  \; .
\label{crit1}
\end{equation}

\noindent
This means that the differential rotation parameter $|q|$ has to be large enough
to be able to generate the magnetic field.  In the numerical examples in Sect.6,
 the condition is not satisfied in the very outerlayers, which have a too weak
 differential rotation.

\begin{figure}[t]
  \resizebox{\hsize}{!}{\includegraphics{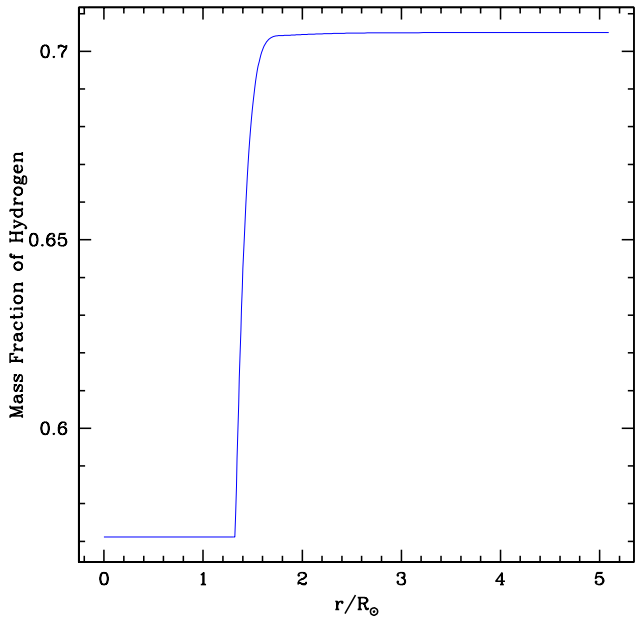}}
  \caption{Internal H--profile in the 15 M$_{\odot}$ test model with $v_{\mathrm{ini}}=300$ km/s.
  The model is 
  at an age of $4.0  \cdot 10^6$ yr. This is the reference model in which 
  we examine the  properties of the magnetic field  in detail.} 
  \label{fx}
\end{figure}

\begin{figure}[t]
  \resizebox{\hsize}{!}{\includegraphics{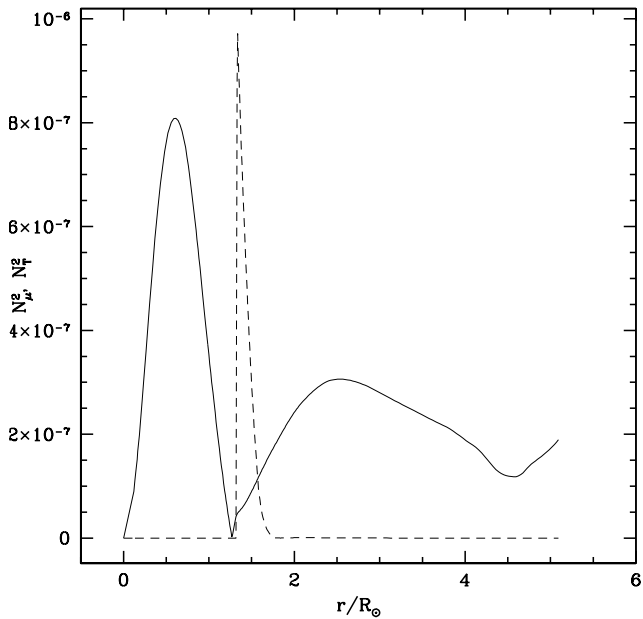}}
  \caption{The oscillation frequencies in the model of Fig.~\ref{fx}.
  $N^2_{\mu}$ is indicated by a continuous line and $N^2_{\mathrm{T}}$ is given
  by a dashed line. }
  \label{fn}
\end{figure}

\subsection{Collection of formulae and recipes}

Let us collect here the various expressions we have for the
diffusion coefficients by the Tayler--Spruit dynamo in radiative zones.
 Case 0 applies when $N_{\mu} > N_{\mathrm{T}}$. Magnetic field is
 present only when the criterion given by Eq.~(\ref{crit0}) is satisfied. Then
the diffusion coefficients for the transport of the angular momentum 
and chemical elements are respectively,

\begin{equation}
 D_{\rm ang0}=r^2 \Omega q^2 \left( {\Omega \over N_{\mu}}\right)^4  \; ,
 \label{Dang0}
\end{equation}

\begin{equation}
 D_{\rm chem0} =r^2 \Omega q^4 \left( {\Omega \over N_{\mu}}\right)^6  \; .
\end{equation}

\noindent
Case 1 applies when $N_{\mu} < N_{\mathrm{T}}$, then account is given to the thermal
diffusivity. Magnetic field is
 present only when the criterion given by Eq.~(\ref{crit1}) is satisfied.
 The diffusion coefficients for the angular
momentum and chemical elements are respectively 
\begin{equation}
 D_{\rm ang1}=r^2 \Omega \left( {\Omega \over N_{\mathrm{T}}}\right)^{1/2} 
 \left({K \over r^2 N_{\mathrm{T}}} \right)^{1/2}  \; ,
 \label{Dang1}
\end{equation}

\begin{equation}
 D_{\rm chem1}=r^2 \Omega \; |q| \left( {\Omega \over N_{\mathrm{T}}}\right)^{3/4} 
 \left({K \over r^2 N_{\mathrm{T}}} \right)^{3/4}    \; .
\end{equation}

\noindent
Normally, these coefficients are larger than those which would be obtained in case 0
 with $N_{\mathrm{T}}$ instead of $N_{\mu}$, because
 the account for thermal effects reduces the buoyancy force which opposes to the
 magnetic instability.  However, as noted by Spruit (\cite{Spruit02}), it may happen 
 in some cases that these
 coefficients with index ``1'' are smaller.  As noted by Spruit (\cite{Spruit02}),
 this is an artefact from the simplification introduced by
 considering only the 2 limiting cases 0 and 1. Spruit
 suggests  to introduce an interpolation formula depending on $q$. We hesitate to do so, 
 because this extra--dependence on the differential rotation parameter $q$  is unphysical,
 since the interpolation should rather depend on the thermal and magnetic  diffusivities $K$
 and $\eta$. Thus, the suggested treatment might
 introduce spurious effects in some evolutionary stages. The general case where thermal 
 effects and $\mu$--gradient are accounted for needs to be worked out
 in future. For now, we prefer to do the following by considering
 the coefficients 

\begin{equation}
 D_{\rm ang1P}=r^2 \Omega q^2 \left( {\Omega \over N_{\mathrm{T}}}\right)^4  \; ,
 \label{Dang1P}
\end{equation}

\begin{equation}
 D_{\rm chem1P}=r^2 \Omega q^4 \left( {\Omega \over N_{\mathrm{T}}}\right)^6  \; .
\end{equation}

\noindent
These are the same equations as in case 0, but with $N_{\mathrm{T}}$ instead of $N_{\mu}$.
This means that we are considering only the restoring force of the thermal gradient
and that we are ignoring the non--adiabatic radiative losses.

For case 1, we compare the coefficient with indices ``1'' and ``1P''
and must take the larger ones. From Fig.~(\ref{fd2}) below, we see that the coefficients 
``1P'' may  be larger than the coefficients  ``1'' 
in some parts of the star. Of course, we have also to test whether 
the magnetic field can be created from differential rotation. For such zones of case 1P, the
criterion for the existence of the magnetic field is evidently  the following one

\begin{equation}
\frac{\Omega}{N_{\mathrm{T}}} \; < \; \frac{1}{\sqrt{3}} = 0.5774 \; ,
\label{crit1P}
\end{equation}
\noindent
and the appropriate test has to be made in the  concerned layers.

\begin{figure}[t]
  \resizebox{\hsize}{!}{\includegraphics{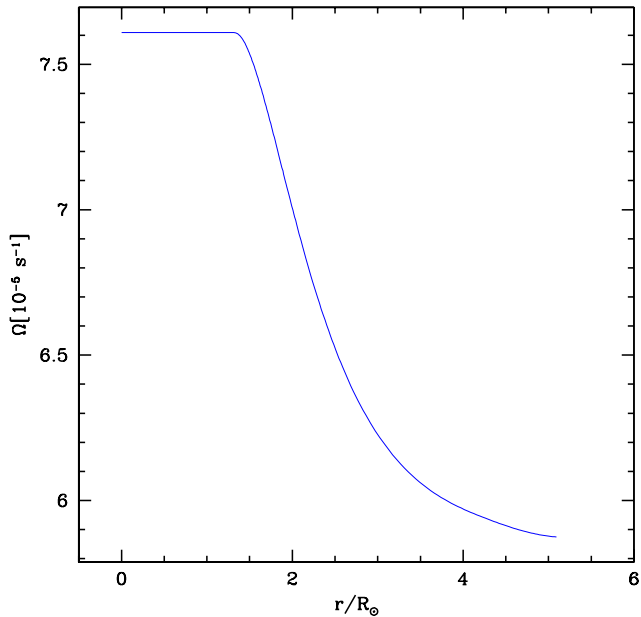}}
  \caption{The distribution of the angular velocity in the reference model of Fig.1 
  with rotation (continuous line), 
  } 
  \label{fom}
\end{figure}

\begin{figure}[t]
  \resizebox{\hsize}{!}{\includegraphics{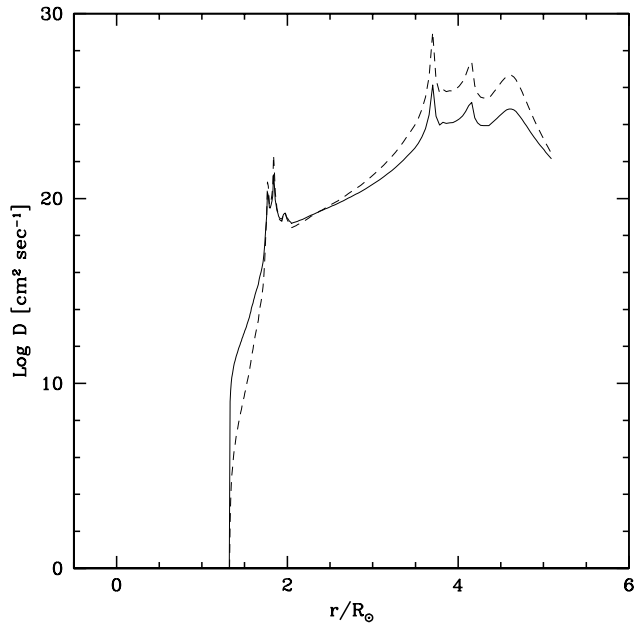}}
  \caption{The  diffusion coefficients $D_{\rm ang0}$ (continuous line) and $D_{\rm chem0}$
  (dashed line)corresponding to case 0.  As discussed in the text, 
  these coefficients apply only in the rising
part between $r/{\mathrm{R_{\odot}}}= 1.32$ and 1.59.} 
  \label{fd}
\end{figure}

\section{Numerical applications to stellar models}

\subsection{Magnetic vs. rotational transports
 of angular momentum and of chemical elements
}

We calculate evolutionary models of a 15 M$_{\odot}$ star with initial composition
$X=0.705$ and $Z=0.02$. The physics of the models (opacities, nuclear reactions, mass loss rates,
treatment of rotation, increase of the mass loss rates with rotation, etc...) 
are  the same as in Meynet \& Maeder (\cite{MMX}). 
We compute the MS evolution of   a model with
an initial velocity of 300 km$\cdot$s$^{-1}$, which leads to an average velocity during the MS phase
of about 220 km$\cdot$s$^{-1}$, which corresponds to the observed average rotation velocity.
We now consider in detail the properties of a particular model with rotation to see 
the various  coefficients and criteria characterizing the growth
of the magnetic field. 
We take the model  at an age $4.17 \cdot 10^6$ yr with a central 
H--content $X_{\mathrm{c}}= 0.571$. The H--profile inside the star is illustrated in 
Fig.~\ref{fx} and the oscillation frequencies $N^2_{\mu}$ and $N^2_{\mathrm{T}}$
in Fig.~\ref{fn}. We see that case 0 applies between 
$r/{\mathrm{R_{\odot}}}= 1.32$ and 1.59, which corresponds to mass coordinates 
5.53  and 7.70 M$_\odot$ respectively. The internal profile $\Omega(r)$ is illustrated 
in Fig.~\ref{fom}.

The  diffusion coefficients $D_{\rm ang0}$  and $D_{\rm chem0}$ are illustrated 
in Fig.~\ref{fd}. These  coefficients grow very fastly at the edge of the core 
since $N^2_{\mu}$ decreases very fastly there  and the dependence in the ratio
$(\frac{\Omega}{N_{\mu}})$ goes like the power 4 and 6 for the two coefficients 
respectively. These coefficients apply only in the rising
part between $r/{\mathrm{R_{\odot}}}= 1.32$ and 1.59 as indicated. Above this value, the main 
restoring force is no longer the $\mu$--gradient, but the stable temperature gradient.

\begin{figure}[t]
  \resizebox{\hsize}{!}{\includegraphics{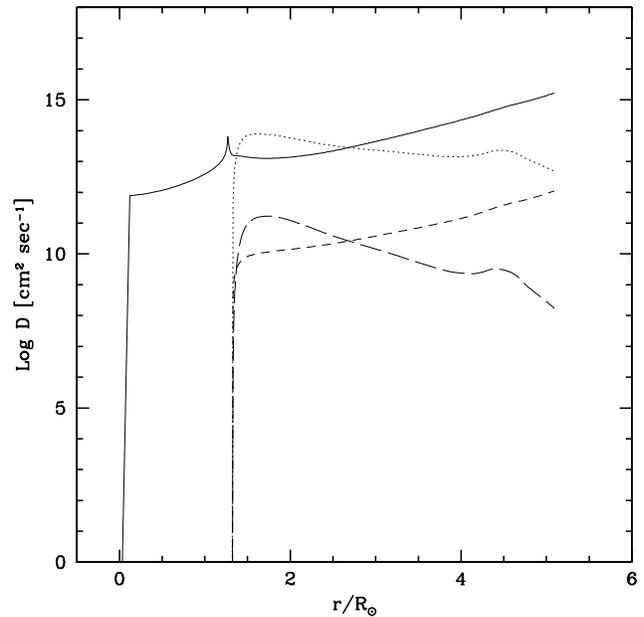}}
  \caption{The diffusion coefficients for angular momentum $D_{\rm ang1}$ (continuous line) 
  and $D_{\rm ang1P}$ (dotted line). The diffusion coefficients for chemical
  elements $D_{\rm chem1}$  (dashed line)  and  $D_{\rm chem1P}$ (long dashed line).
  In each case, the largest diffusion coefficient has to be taken.}
  \label{fd2}
\end{figure}

In case 1, where thermal gradients dominate, the diffusion coefficients are
illustrated in Fig.~\ref{fd2}. We see that  for the transports of 
angular momentum and of  chemical elements, the coefficients with indices ``1''
dominate in the external parts of the star above $r/R_{\odot}= 2.69$.
 However, we also notice that in a sizeable
region, i.e. between  $r/R_{\odot}=1.59$ and $2.69$, 
the coefficients with indices   ``${1{\mathrm{P}}}$'' dominate
over those with  ``${1}$''. This occurs, as expected, at some limited 
distance above the edge of  the convective core, at the place where the $\mu$--gradient
becomes small enough, but is still different from zero. In these regions, more general developments
having case 0 and 1 as limiting cases would be a progress.  We see that the differences between
the two cases ``${1{\mathrm{P}}}$''  and  ``${1}$'' amounts to a maximum of 0.4 dex and 0.6 dex 
for the transport of angular momentum and chemical elements respectively. This is limited, but
non negligible, and it may justify a further study of the physics of the 
general case. However,  we  notice that this difference is  small when compared
to the differences resulting from the inclusion of the magnetic field
or not, which as  shown below amounts to several orders of magnitude.
Thus, we conclude that the present coefficients of diffusion 
need to be further improved, but they nevertheless describe correctly the 
main results of the inclusion of the magnetic field.

\begin{table*}
\caption{Structural parameters of the model of 15 M$_{\odot}$ with
$v_{\mathrm{ini}}= 300$ km/s when $X_{\mathrm{c}}= 0.571$} \label{tbl-1}
\begin{center}\scriptsize
\begin{tabular}{ccccccccccc}
\hline
    &         &         &         &         &     &      &      &      &    &      \\
$M_r/M_\odot$  & $r/{\mathrm{R_{\odot}}}$ & $\Omega$ & ${d\ln\Omega \over d\ln r}$ &  $\nabla_\mu$ & $\nabla_{\rm rad}$ & $\nabla_{\rm ad}$ & $g$ & $\delta$ & $H_p$  & $K$   \\
    &         &         &         &         &     &      &      &      &    &      \\
6.5726 &  1.445 &  0.76E-04 & -0.11E+00 & 0.18E+00 & 0.317 & 0.337 & 0.86E+05 & 1.334 & 0.33E+11 & 0.35E+10\\
    &         &         &         &         &     &      &      &      &   &       \\
11.0386 & 2.080 & 0.69E-04 & -0.33E+00 & 0.32E-03 & 0.269 & 0.348 & 0.69E+05 & 1.255 & 0.27E+11 & 0.79E+10\\
    &         &         &         &         &     &      &      &      &   &       \\
14.9421 & 4.159 & 0.60E-04 & -0.84E-01 & 0.14E-06 & 0.225 & 0.326 & 0.23E+05 & 1.455 & 0.19E+11 & 0.81E+12\\
    &         &         &         &         &     &      &      &      &    &      \\
\end{tabular}
\end{center}
\end{table*}

\begin{table*}
\caption{Diffusion coefficients  of the model of 15 M$_{\odot}$ with
$v_{\mathrm{ini}}= 300$ km/s when $X_{\mathrm{c}}= 0.571$ } \label{tbl-2}
\begin{center}\scriptsize
\begin{tabular}{cccccccccccc}
\hline
    &         &         &         &         &     &      &      &      &    &    &          \\
$M_r/M_\odot$  &  $N_\mu^2$ & $N_T^2$ &  $D_{\rm ang0}$ & $D_{\rm chem0}$ & $D_{\rm ang1}$ & $D_{\rm chem1}$ & $D_{\rm ang1P}$ & $D_{\rm chem1P}$  & $D_{\rm shear}$ &  $U_{\rm
circ}$   \\
    &         &         &         &         &     &      &      &      &    &    &          \\
6.5726 &   0.48E-06 & 0.73E-07  &
   0.12E+13 & 0.17E+09 &  0.15E+14&0.68E+10  & 0.55E+14 & 0.49E+11  &
   0.40E+04 &    -0.19E-03 \\
 $U_{\rm magn}$     &         &         &         & -0.66E+01  &  & -0.77E+02  & &  -0.29E+03 &  &  &   \\
      &         &         &         &         &     &      &      &      &    &    &          \\
11.0386 &  0.84E-09 & 0.26E-06  &
  0.51E+19 & 0.31E+19 & 0.14E+14 & 0.15E+11 & 0.52E+14 & 0.10E+12 & 
  0.28E+07 &   -0.23E-04\\
  $U_{\rm magn}$    &          &         &         & -0.57E+08  &  &-0.16E+03  & & -0.59E+03&  &  &   \\
    &         &         &         &         &     &      &      &      &    &    &          \\
14.9421 &  0.17E-12 & 0.17E-06 & 
  0.16E+26 & 0.24E+28 &0.29E+15  & 0.19E+12 & 0.15E+14 & 0.23E+10  &
  0.19E+09 &    0.13E-03 \\
    $U_{\rm magn}$    &          &         &         & -0.23E+14 &  &  &-0.42E+03 & -0.23E+02&  &  &   \\
    &         &         &         &         &     &      &      &      &    &    &          \\

\end{tabular}
\end{center}

\end{table*}

\begin{figure}[t]
  \resizebox{\hsize}{!}{\includegraphics{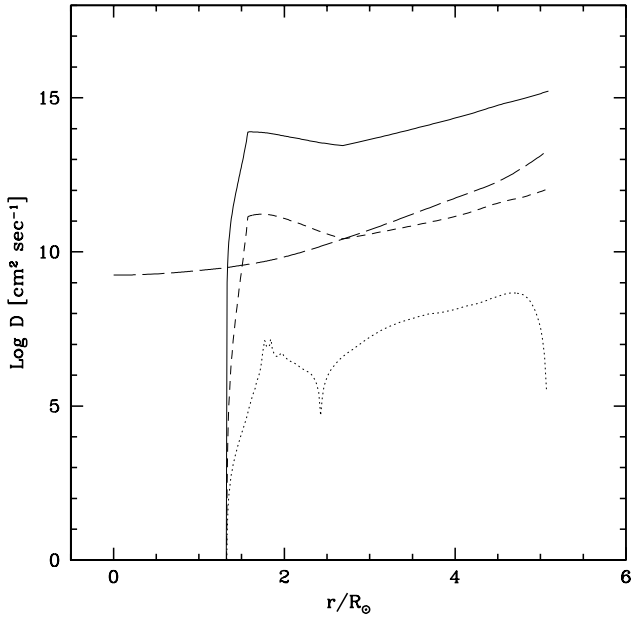}}
  \caption{The figure shows the complete description
  of the diffusion coefficient $D_{\mathrm{ang}}$ for angular momentum (continuous line) made according to 
  prescriptions of Sect.~(5.2).  The complete description of the diffusion coefficient
  $D_{\mathrm{chem}}$  for chemical elements is given by the dashed line.
  The dotted line shows the shear diffusion coefficient $D_{\mathrm{shear}}$
  from Maeder (\cite{MII}). The thermal diffusivity coefficient $K$ is represented
  by a long--dashed line.  }
  \label{fdi}
\end{figure}

Tables \ref{tbl-1} and \ref{tbl-2} provides some useful structural parameters, the
diffusion coefficients and the velocities of meridional circulation and of  magnetic
instability at 3 locations 
in the reference model of 15 M$_{\odot}$  with
an initial velocity of 300 km s$^{-1}$ at an age $4.17 \cdot 10^6$ yr. 
The three levels considered  illustrate the  case 0, 
1P and 1 respectively. These Tables permit further quantitative analysis
 of the various terms  intervening in the  equations.

Fig.~\ref{fdi} shows the comparison of the diffusion coefficients due
to the magnetic field compared to the diffusion coefficient $D_{\mathrm{shear}}$ by shear
instability in the  rotating star and to the thermal diffusivity $K$.
The transport of angular momentum by the
magnetic field is 6--7 orders of magnitude stronger than by shear instability in a rotating star.
Similarly, as discussed in Sect.~(4.2), the  magnetic  transport of angular momentum
is also 4--7 orders of magnitude larger than by meridional circulation. Therefore, we conclude
that transport of angular momentum by magnetic field is totally dominating, if magnetic field is present.

For the transport of chemical elements, the difference between the two diffusion coefficients
amount to 3--5 orders of magnitude in favour of the transport by magnetic field. The
difference is especially large in deep regions at some distance of 
the convective core. At the very edge of the convective core, 
the dependence of the coefficient $D_{\mathrm{chem, 0}}$ 
in the power 6 of $(\frac{\Omega}{N_{\mu}})$ reduces the magnetic diffusion
drastically, so that as mentioned in Sect.~4.2 the ratio
of the transport of chemical elements by the magnetic instability to the transport by 
circulation may amount to about 1 order of magnitude.
 Some tests indicate that this  makes  the chemical enrichments in helium and nitrogen at the
stellar surface are  stronger, but not too different,
 from those without magnetic field, despite the fact that
the diffusion coefficients with magnetic field are orders 
of magnitude larger over most of the stellar interior.
On the whole, we see that for chemical mixing also, the magnetic 
instability plays a great role.

It is also interesting to see that diffusion coefficients by the magnetic field
are almost equal (transport of chemical elements) or even larger
 (transport of the angular momentum)  than  the thermal diffusivity.
 This means that magnetic effects by the Tayler--Spruit dynamo
 are in general equal or larger than thermal effects. Globally thermal
  effects may be relatively more significant in the outer layers.

\begin{figure}[t]
 \resizebox{\hsize}{!}{\includegraphics{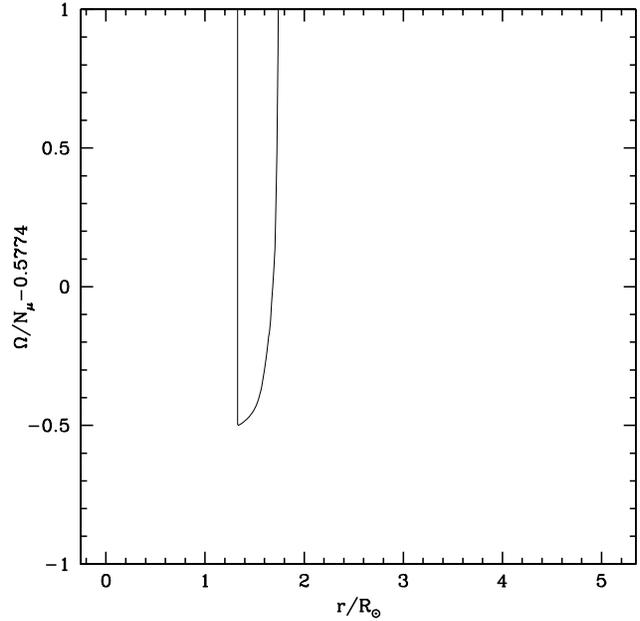}}
  \caption{
  The difference  $\frac{\Omega}{N_{\mu}} - \frac{1}{\sqrt{3}}$ when  case 0 applies,
  i.e. in the region  between $r/{\mathrm{R_{\odot}}}= 1.32 $ and $1.59$.  
  According to criterion (\ref{crit0}), if this difference is negative,
   magnetic field is created. Thus, we see that in the region where the $\mu$--gradient
   dominates, magnetic field can be generated by the Tayler--Spruit dynamo.}
  \label{onmu}
\end{figure}

\subsection{Existence of a magnetic field due to the energy
condition of differential rotation}

\begin{figure}[t]
  \resizebox{\hsize}{!}{\includegraphics{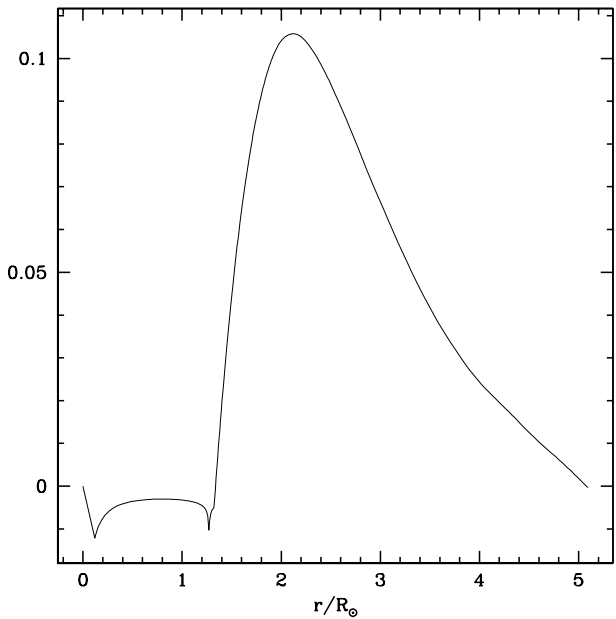}}
  \caption{The quantity  $\frac{|q|}{3} - \left(\frac{\Omega}{N_{\mathrm{T}}}\right)^{\frac{1}{4}}
\left(\frac{K}{r^2 \; N_{\mathrm{T}}}\right)^{\frac{1}{4}}$ given by Eq.~(\ref{crit1}) vs. $r/{\mathrm{R_{\odot}}}$.
When this quantity is positive in the region where case 1 applies, i.e. 
 above 
$r/{\mathrm{R_{\odot}}}= 2.69 $, differential rotation is sufficient to generate the magnetic 
field, which is the case here.
 } 
  \label{ontt}
\end{figure}

An important question in the models is to determine at each shell mass
whether differential rotation described by parameter $q$
is sufficient to create the Tayler--Spruit dynamo
 to produce a magnetic field. We examine here whether these conditions are 
 fulfilled in the model with rotation only studied in the previous 
 subsection.  Firstly, we examine the regions adjacent
 to the core   between $r/{\mathrm{R_{\odot}}}= 1.32$ and 1.59,
  where case 0 applies, since $N_{\mu}$ dominates. 
 Fig.~\ref{onmu} shows the difference $\frac{\Omega}{N_{\mu}} - \frac{1}{\sqrt{3}}$,
 we see that in the concerned region between $r/{\mathrm{R_{\odot}}}= 1.32 $ and $1.59$,
 this difference is negative, thus magnetic field is present there. This is an interesting result
 because it means that despite the strong restoring buoyancy force due to the
 very large $\mu$--gradient, the differential rotation is high enough to
 develop magnetic instability.
 At each time  step during evolution, such tests need to be performed.
 
 Secondly, we examine the zone above  $r/{\mathrm{R_{\odot}}}= 1.59$ and up to
 $r/{\mathrm{R_{\odot}}}= 2.69$, where case 1P applies. There, we check for the difference
 $\frac{\Omega}{N_{\mathrm{T}}} - \frac{1}{\sqrt{3}}$. If this expression is negative,
  magnetic field is created. This expression lies   between
 -0.40 and -0.50 in this whole intermediate region. Thus, we conclude that
 magnetic field is also present there. 
 
 The criterion for the existence of the magnetic field in the external zone,
 which corresponds to  case 1 is given by Eq.~(\ref{crit1}). From Fig.~(\ref{ontt}),
 we see that in this zone 
 which lies between $r/{\mathrm{R_{\odot}}}= 2.70$ and the surface, the difference 
 $\frac{|q|}{3} \; > \; \left(\frac{\Omega}{N_{\mathrm{T}}}\right)^{\frac{1}{4}}
\left(\frac{K}{r^2 \; N_{\mathrm{T}}}\right)^{\frac{1}{4}}$ is generally
positive, which means that magnetic field is present over that region. However,
we notice that very close to the surface this difference goes to zero. This means
that at the surface, differential rotation is becoming insufficient to generate the
magnetic instability. This is interesting because it may explain why there is in general no 
strong magnetic field observed at the surface of OB stars (cf. Mathys \cite{Mathys03}), despite 
the likely existence of a strong internal field created the Tayler--Spruit instability.

These results  show that in a rotating star,
 the conditions on the differential rotation for the growth of
  the magnetic field are largely realized, with the
 possible exception of the superficial layers. 
 
 \section{Conclusions}
 
 The main conclusion is that the Tayler instability and the Tayler--Spruit dynamo 
 are of major
 importance for stellar evolution, both for the transport of angular momentum and
 for the transport of chemical elements. Future evolutionary models applying 
 the results of this work will be made to study
  the results on tracks, surface composition, rotation, etc...

 It is likely that  in a rotating star calculated with magnetic field from the beginning,
 the differential rotation is very much reduced by the magnetic transport
 of angular momentum described above. We may suspect that differential
 rotation will be reduced down to a stage where the criterion (\ref{critgen}) discussed
 in Sect. 5.1 is just marginally satisfied, i.e 
 \begin{equation}
 |q| = \sqrt{3} \; \; \frac{\omega_{\mathrm{A}}}{\Omega}  \; .
 \label{limit}
 \end{equation} 

\noindent
Indeed, if differential rotation is higher than given by this criterion, magnetic field
develops and the associated coupling reduces differential rotation. 
If, at the opposite, differential rotation is lower 
than given by criterion (\ref{limit}), the growth of mean molecular weight by nuclear reactions in the central regions together with angular momentum conservation will produce an
enhancement of differential rotation. Thus, a stage of marginal equilibrium 
is most likely reached during MS evolution. It may also be that the outer layers
never have sufficient differential rotation to build Tayler-Spruit dynamo.
Further numerical models will explore the evolution of stars with rotation and magnetic field, and
analyse the coupling between  magnetic field and differential rotation.

 
  

\begin{acknowledgements}
We express our best thanks to Dr. H.C. Spruit for very useful comments.
The most valuable  remarks of an unknown referee are also acknowledged with thanks.
\end{acknowledgements}

\end{document}